\documentclass[conference]{IEEEtran}
\IEEEoverridecommandlockouts
\usepackage{cite}
\usepackage{amsmath,amssymb,amsfonts}
\usepackage{algorithmic}
\usepackage{graphicx}
\usepackage{textcomp}
\usepackage{xcolor}
\usepackage{placeins}

\def\BibTeX{{\rm B\kern-.05em{\sc i\kern-.025em b}\kern-.08em
    T\kern-.1667em\lower.7ex\hbox{E}\kern-.125emX}}
\begin{document}

\title{Weather-Driven Priority Charging for Battery Storage Systems in Hybrid Renewable Energy Grids \\
}

\author{\IEEEauthorblockN{Dhrumil Bhatt\textsuperscript{*}}
\IEEEauthorblockA{\textit{Department of Electrical and
} \\
\textit{Electronics Engineering}\\
\textit{Manipal Institute of Technology}\\
\textit{Manipal Academy of Higher Education}\\
Manipal, India \\
dhrumil.bhatt@gmail.com}
\and
\IEEEauthorblockN{Siddharth Penumatsa\textsuperscript{*}}
\IEEEauthorblockA{\textit{Department of Electronics and} \\
\textit{Communications Engineering}\\
\textit{Manipal Institute of Technology}\\
\textit{Manipal Academy of Higher Education}\\
Manipal, India \\
psiddharthv06@gmail.com}
\and
\IEEEauthorblockN{Nirbhay Singhal\textsuperscript{†}}
\IEEEauthorblockA{\textit{Department of Electrical and} \\
\textit{Electronics Engineering}\\
\textit{Manipal Institute of Technology}\\
\textit{Manipal Academy of Higher Education}\\
Manipal, India \\
nirbhaysinghal09@gmail.com}
\textit{\textsuperscript{†}Corresponding Author}

}

\maketitle

\begin{abstract}
The integration of renewable energy into the power grid is often hindered by its fragmented infrastructure, leading to inefficient utilization due to the variability of energy production and its reliance on weather conditions. Battery storage systems, while essential for stabilizing energy supply, face challenges like suboptimal energy distribution, accelerating battery degradation, and reducing operational efficiency. This paper presents a novel solution to these challenges by developing a large-scale, interconnected renewable energy network that optimizes energy storage and distribution. The proposed system includes strategically placed battery storage facilities that stabilize energy production by compensating for fluctuations in renewable output. A priority charging algorithm, informed by real-time weather forecasting and load monitoring, ensures that the most suitable battery systems are charged under varying conditions. Within each storage facility, a secondary priority charging algorithm minimizes battery degradation by ranking batteries based on critical parameters such as state of health (SoH) and state of charge (SoC) and deciding which to charge. This comprehensive approach enhances the efficiency and longevity of battery storage systems, offering a more reliable and resilient renewable energy infrastructure.
\end{abstract}
\footnote{\textsuperscript{*}Authors contributed equally to this work.}

\begin{IEEEkeywords}
Renewable energy, Priority charging, Battery storage systems, Weather forecasting, Load monitoring, Hybrid energy systems, Grid stability, Energy optimization, Smart grid management
\end{IEEEkeywords}

\section{Introduction}
As global fossil fuel reserves continue to diminish and energy demands increase, integrating renewable energy sources has become critical in addressing the challenges posed by climate change. During the transitional phase from fossil fuel dependency to fully renewable grids, we face significant challenges, mainly when renewable energy production is not at its peak.

Solar power generation is subject to diurnal cycles and weather conditions, leading to fluctuating output throughout the day. For instance, solar irradiance can vary between 0 and 1,000 W/m² depending on cloud cover and geographic location. On a clear day, solar panels can operate near their peak capacity, but during cloudy conditions, the output can decrease by as much as 50-90\%\ depending on cloud density \cite{Bird2013}. This variability results in supply-demand mismatches and can strain grid management systems. 

Wind power similarly faces significant challenges due to the unpredictable nature of wind speeds. Wind turbines are typically designed to operate within a specific range of wind speeds, with cut-in speeds (the minimum wind speed needed to generate electricity) around 3 to 4 m/s and cut-out speeds (the maximum safe operating speed) around 25 m/s. Outside this range, turbines either produce no power or must be shut down to prevent damage. The variability of wind speeds limits generation capacity and can lead to rapid changes in power output, destabilizing grid systems. These fluctuations make it challenging to ensure a consistent supply of electricity from wind power. \cite{Hassan2023}

Overcoming fluctuations in energy for the grid associated with renewable energy integration requires a strategic push towards hybrid energy systems, which mitigate reliance on any singular renewable source. Combining several renewable sources to work together in different locations or types of generation will significantly improve the grid's stability. The work proposes an infrastructure consisting of sizeable modular battery storage systems, each composed of multiple battery units, and each unit is managed by state-of-the-art battery management systems (BMS) for its internal cells. These systems will ensure optimal efficiency at the individual cell level, allowing future improvements to focus on increasing overall system efficiency. These systems will be placed across the region uniformly rather than next to the energy sources to act as a common energy source. Traditional infrastructure often places energy storage close to the generation source (e.g., wind farms or solar arrays). The proposed approach of distributing storage uniformly across regions helps reduce the risk of localized energy shortages and allows for better balancing of supply and demand across larger areas.
A priority charging algorithm will be implemented for these storage systems, utilizing their state of charge levels in conjunction with weather forecasting for predicted power output and load monitoring for predicted power usage. 

A secondary priority-based charging algorithm will be developed to optimize energy distribution among individual battery units. This algorithm will leverage each unit’s state of charge (SoC) and state of health (SoH) values, assigning a rank based on their respective parameters. Current energy storage systems often treat batteries uniformly, accelerating degradation and necessitating more frequent replacements. In contrast, the proposed algorithm aims to mitigate wear on batteries in poorer conditions by prioritizing healthier units, thereby enhancing their usage and extending the overall lifespan of the battery system.

\section{Literature Review }
\subsection{Integrated Systems: Combining Weather and Load Forecasting}
Weather forecasting has been critical for renewable energy grid management, particularly solar and wind power forecasting. Accurate weather predictions allow for anticipating periods of high renewable energy production (e.g., sunny or windy conditions) and can inform the timing of battery charging. Many studies have investigated different forecasting techniques—statistical, machine learning, and hybrid methods—to improve the accuracy of renewable energy generation predictions\cite{Ying2023,Singla2022,Giebel2017}. However, there is a lack of research integrating exact weather forecasting models directly with battery storage control algorithms. Most work focuses on improving forecast accuracy, with limited attention to how these forecasts can be leveraged in real-time decision-making for battery management systems.

Effective battery management for grid stability also requires accurate load forecasting. Load forecasting models predict future electricity consumption based on historical data and external factors, such as time of day, temperature, and economic activity. This enables utilities to optimize battery discharge during peak demand periods. While advances in machine learning have significantly improved load forecasting accuracy, few studies have examined how load forecasts could be used with weather forecasts to create a more holistic, optimized approach to battery storage management\cite{Amjady2001}.

Although considerable research exists on weather and load forecasting as independent areas, there is a significant gap in combining these two forecasting tools for battery storage management in renewable grids. Some studies have explored hybrid systems integrating weather and load data to improve grid stability. However, these systems often lack real-time implementation and dynamic priority charging algorithms that adjust based on the immediate needs of the grid, such as sudden weather changes or unexpected shifts in demand\cite{Albogamy2022}.

\subsection{Priority Charging}
Electric vehicle (EV) charging infrastructures have increasingly adopted intelligent charging algorithms emphasizing real-time, dynamic priority allocation, optimizing vehicle charging times based on grid demand and EV user requirements. Some systems consider immediate demand and urgency in-vehicle charging.
These approaches ensure that EVs receive priority charging access, which addresses user convenience and the operational requirements of transportation systems\cite{Zheng2016,Xu2016}. However, similar work has not been done on priority-based charging systems for renewable energy grid storage. Battery storage systems for renewable energy often operate on longer timescales, with less real-time priority, but sophisticated priority charging algorithms could optimize energy storage efficiency.

\subsection{Challenges and Gaps in Research}
While advancements have been made in battery management systems, weather forecasting, and load forecasting, several key research gaps persist:
\begin{itemize}
    \item \textbf{Lack of integrated models:} Few studies have integrated weather and load forecasting into a single, real-time framework for battery storage management in renewable grids.
    \item \textbf{Limited research on real-time adaptability:} Most optimization models use static assumptions and do not fully exploit real-time weather and load data to adjust battery storage operations dynamically.
    \item \textbf{Inadequate focus on priority-based charging:} Existing research does not sufficiently address the development of priority charging strategies that dynamically prioritize energy storage based on immediate grid needs.
    
\end{itemize}

\section{Methodology}
\subsection{Stability Driven Priority Algorithm}
The algorithm optimizes grid stability and ensures a continuous power supply, even during sudden changes. It serves as a control mechanism for battery storage systems, using a range of input parameters, including predicted power generation from renewable sources, forecasted power demand based on load monitoring, and the current state of charge (SoC) of individual battery storage systems.

The battery storage system is organized such that each system contains ten individual battery units, with the State of Charge (SoC) of the system calculated as the sum of the energy of its units. The following equation determines the SoC of each storage system:

\begin{equation}
\small
S_{\text{sys}} = \left( \frac{\sum_{i=1}^{n}{E_i}}{B} \right) \times 100
\end{equation}

\noindent
where:
\begin{itemize}
    \item $S_{\text{sys}}$: Storage System SoC (State of Charge) in percentage
    \item $E_i$: Energy of the $i$-th battery unit
    \item $B$: System Capacity
    \item $n$: Number of Units
\end{itemize}

\subsubsection{Priority Charging Logic}

The system prioritizes charging for storage systems based on the connected load demands. For each connected load, the energy demand required to reach the desired charge target is calculated as the connected load along with twenty-five percent of its capacity as a buffer in case of sudden changes in load demands. The total system capacity constrains the charge target, and the required energy to meet the charge target is given by the charge target minus the current system SoC.

If the available energy from the source exceeds the total charge needed, the system is charged up to its maximum capacity. Any remaining energy is redistributed to other storage systems in the grid.

\subsubsection{Load Distribution and Discharge Logic}

The load distribution mechanism ensures proportional energy discharge among the connected systems based on their SoC. The total energy available from the connected systems for each load is calculated as:

\begin{equation}
\small
E_{\text{avail}} = \sum_{i=1}^{n} \left( \frac{S_i}{100} \times C \right)
\end{equation}

\noindent
where:
\begin{itemize}
    \item $E_{\text{avail}}$: Energy Available
    \item $S_i$: State of Charge of the $i$-th storage system (in percentage)
    \item $C$: System Capacity
    \item $n$: Number of Systems
\end{itemize}

The load demand is scaled down proportionally if the total available energy exceeds the demand. Each storage system then contributes to the discharge based on its SoC:

\begin{equation}
\small
C_i = \frac{S_i \times C}{E_{\text{avail}}} \times D
\end{equation}

\noindent
where:
\begin{itemize}
    \item $C_i$: System Contribution
    \item $S_i$: State of Charge of the $i$-th system (in percentage)
    \item $C$: Capacity of the system
    \item $E_{\text{avail}}$: Total Energy Available
    \item $D$: Load Demand
\end{itemize}

The energy contribution from each system is equally distributed among its ten units, and the SoC of each unit is reduced accordingly. This approach ensures balanced discharge from each system while meeting load demands efficiently.

The algorithm efficiently manages energy distribution by integrating predictive weather data, which informs expected renewable output, with real-time load demands. It prioritizes charging the most suitable battery systems based on their capacity to absorb and store energy effectively while preparing for future energy demands. The ultimate goal is to ensure that energy storage systems are adequately charged to meet anticipated needs, providing any of their capacities ever hit zero.

\begin{figure}[htbp]
\centerline{\includegraphics[scale=0.33]{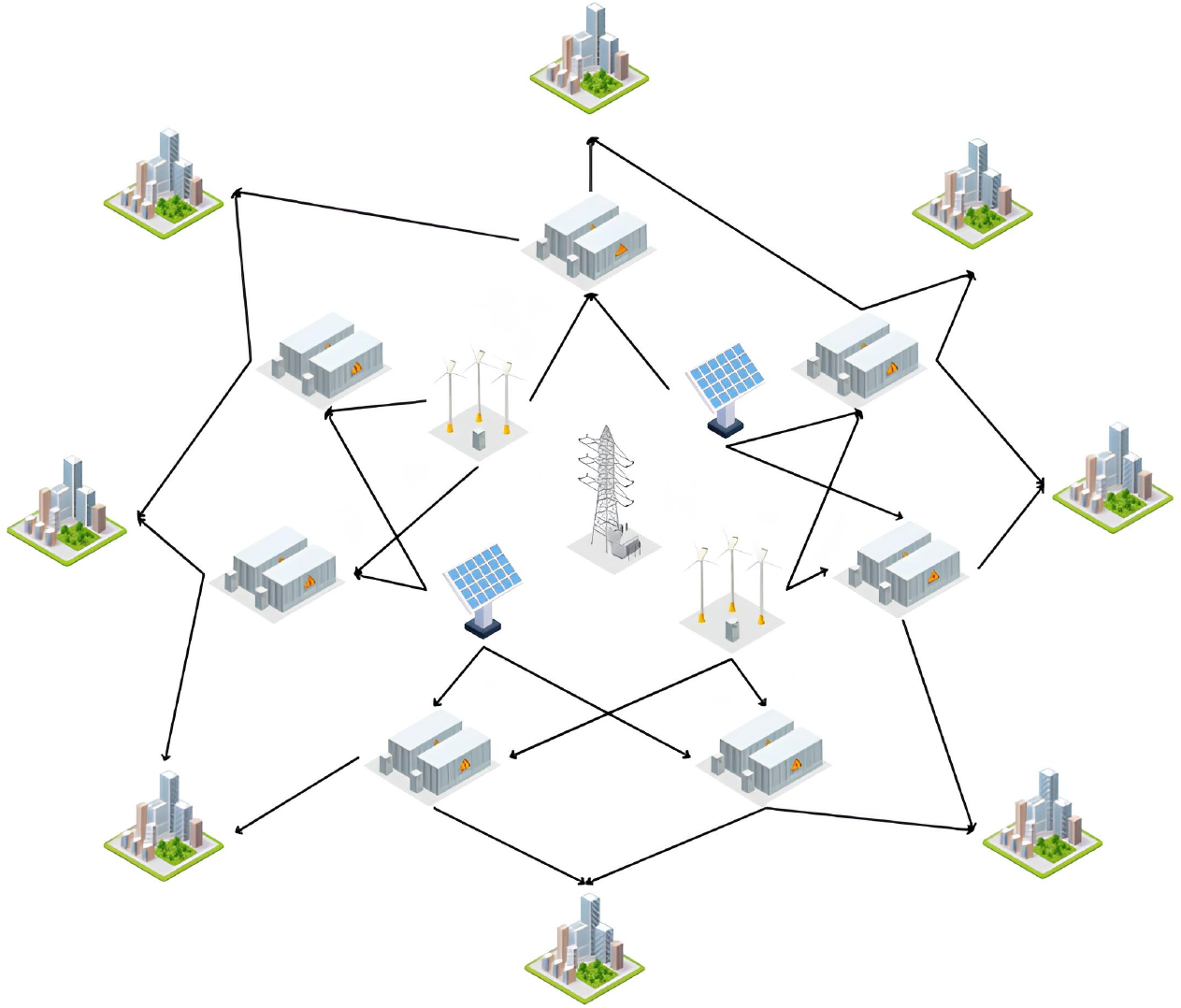}}
\caption{This figure illustrates the central base station wirelessly interconnected with the hybrid energy system, which integrates solar panels and wind turbines as the primary energy generation sources, connected to centralized battery storage units that store excess power for later use and load centers, representing towns or cities.}
\label{fig:hybrid}
\end{figure}

\subsubsection{Weather Forecasting}

The Solcast API provides real-time solar irradiance data and forecasts by modeling atmospheric conditions such as cloud cover and sunlight intensity\cite{Solcast}. The algorithm fetches predicted solar energy output by retrieving Solcast data for two grid solar plants and calculates the expected energy output based on the data, panel efficiency, and system capacity. For example, the predicted power output for a solar plant is derived using\cite{Sawle2016}:

\begin{equation}
P_{\text{solar}} = I_T A_{\text{PV}} \eta_{\text{PV}}
\end{equation}

\noindent
where:
\begin{itemize}
    \item $P_{\text{solar}}$ is the predicted solar power output.
    \item $I_T$ is the total solar irradiance (W/m²) incident on the PV array.
    \item $A_{\text{PV}}$ is the area of the photovoltaic (PV) array (m²).
    \item $\eta_{\text{PV}}$ is the efficiency of the PV array.
\end{itemize}

This prediction is periodically updated to adjust charging priorities.

The same API is employed for wind energy to predict power output based on wind speed forecasts. Wind turbine power output is estimated using the formula\cite{Sawle2016}:

\begin{equation}
P_W = \frac{C_P \rho A V^3}{2}
\end{equation}

\noindent
where:
\begin{itemize}
    \item $P_W$ is the predicted wind power output (W).
    \item $C_P$ is the power coefficient (dimensionless), representing the efficiency of the wind turbine.
    \item $\rho$ is the air density (kg/m³).
    \item $A$ is the frontal area of the wind turbine (m²).
    \item $V$ is the wind speed (m/s).
\end{itemize}

Wind forecasts for the two wind farms are retrieved and used to calculate energy generation. This predicted output helps the algorithm determine how to distribute the power generated and what can be stored in the battery systems.

\subsubsection{Load Monitoring and Prediction}

An essential requirement for ensuring system stability is accurately predicting loads. A thorough literature review of different methods was conducted to explore load demand prediction\cite{Wazirali2023,Amjady2001}. The algorithm has been tested on existing historical load data to ensure accuracy in load predictions. In real-time applications, the stability-driven algorithm will be fed load predictions for all loads connected to the hybrid grid. Based on SoC and expected load, the algorithm will estimate which system should be prioritized for charging to ensure grid stability. SARIMA (Seasonal AutoRegressive Integrated Moving Average) is an extension of ARIMA (AutoRegressive Integrated Moving Average) that incorporates seasonality into time series modeling. In load forecasting, particularly for battery storage systems in renewable energy grids, weather patterns play a significant role. SARIMA is chosen over ARIMA because the solar irradiance and wind patterns change seasonally and the SARIMA is better equipped with prediction of future values with seasonal deviations as compared to ARIMA. \cite{Goswami2020}.

\subsection{Health Efficiency Priority Algorithm}

A separate algorithm has been designed to optimize the health management of individual battery units within larger battery storage systems. The core of this algorithm is a battery ranking method that leverages a battery score calculated by combining the state of health (SoH) and state of charge (SoC).  

Each battery unit's State of Health (SoH) degrades over time due to both charging and discharging processes. The degradation is modeled using predefined constants for charge and discharge degradation rates, applied proportionally to the energy processed by each unit during charging or discharging events.

\begin{equation}
\small
D_{\text{ch}} = \frac{Q_{\text{ch}} \times R_{\text{ch}}}{C}
\end{equation}

\begin{equation}
\small
D_{\text{dis}} = \frac{Q_{\text{dis}} \times R_{\text{dis}}}{C}
\end{equation}

\noindent
where:
\begin{itemize}
    \item $D_{\text{ch}}$: SoH Degradation during charging
    \item $D_{\text{dis}}$: SoH Degradation during discharging
    \item $Q_{\text{ch}}$: Charge Amount
    \item $Q_{\text{dis}}$: Discharge Amount
    \item $R_{\text{ch}}$: Charge Degradation Rate
    \item $R_{\text{dis}}$: Discharge Degradation Rate
    \item $C$: Unit Capacity
\end{itemize}
This degradation logic ensures that the SoH of each unit is dynamically updated based on energy flows, capturing the gradual wear over time. Considering the long-term health of the battery units, this model enables improved battery performance over its lifecycle. 

The degradation rates are selected to reflect realistic operational conditions, allowing for optimized maintenance schedules and ensuring each system operates efficiently within its capacity.

\begin{figure}[htbp]
\centerline{\includegraphics[scale=0.6]{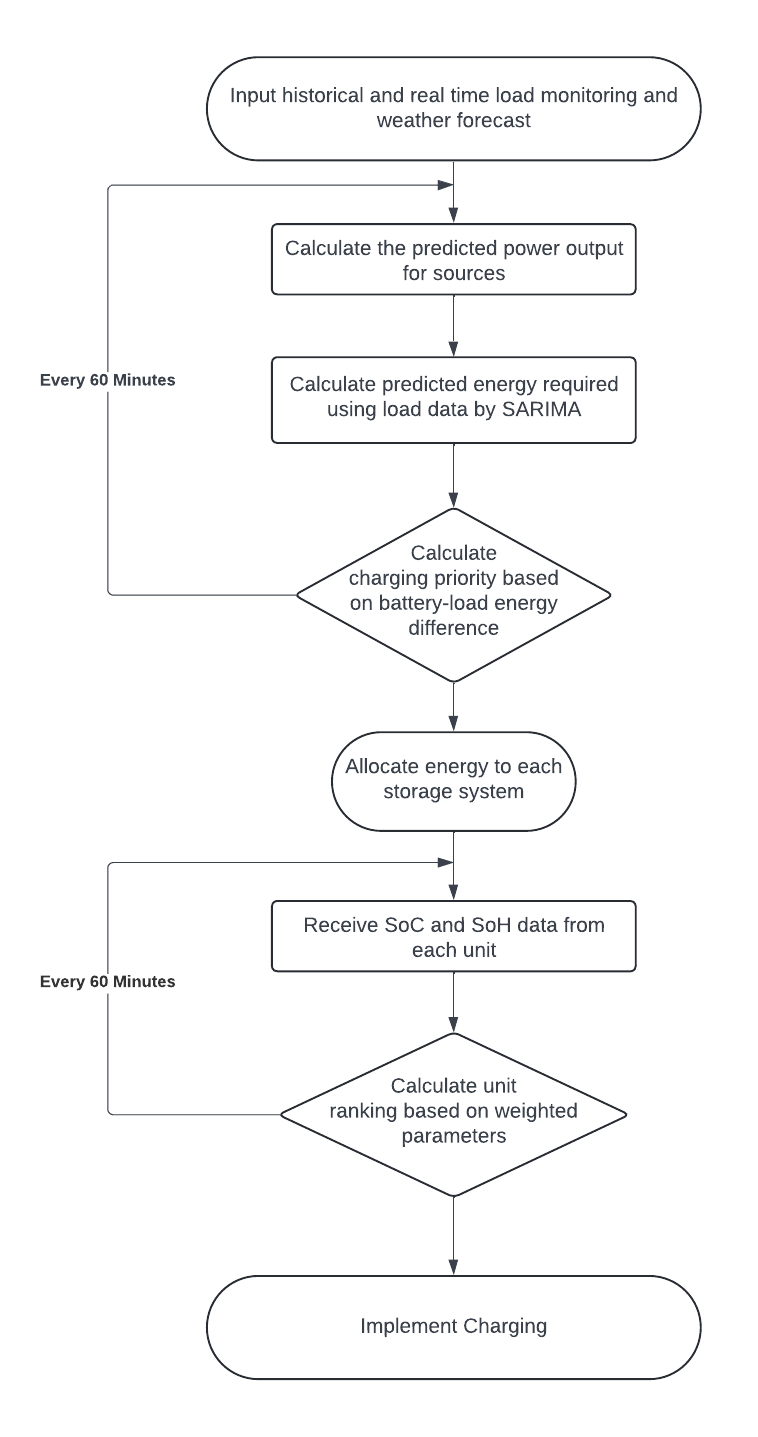}}
\caption{Given above is the logic flow for the combined priority charging and health efficiency algorithm}
\label{fig:flowchart}
\end{figure}
\section{Experimental Setup}

This hybrid energy system integrates renewable energy sources, battery storage systems, and multiple loads connected through a central grid. The entire system has been designed, simulated, and tested on C. All values were converted to work on a daily basis and each algorithm was tested several times with different values to verify system integrity.
In our experimental setup, the load-storage system connectivity is defined as follows: 

Load 0 is connected to Storage Systems 1, 2, and 3;

Load 1 is connected to Storage Systems 2, 3, and 4;

Load 2 is connected to Storage Systems 3, 4, and 5;

Load 3 is connected to Storage Systems 4, 5, and 6; 

Load 4 is connected to Storage Systems 5, 6, and 7;

Load 5 is connected to Storage Systems 1, 6, and 7;

Load 6 is connected to Storage Systems 1, 2, and 7;

Load 7 is connected to Storage Systems 2, 3, and 7.

Systems 1 to 4 are connected to the first two sources, and Systems 4 to 7 are connected to the other two sources. System 4 is connected to all four renewable energy sources. The State of Health and the State of Charge have been assigned weights of 0.5 each to rank units.

Energy has been showcased in MW(Megawatts) for simplicity.

\subsection{System Components}

\subsubsection{Renewable Energy Sources}
\begin{itemize}
    \item \textbf{Wind Power Plants (2 Units):} Two wind plants generate power based on wind speed and atmospheric conditions. Both have been assumed to have average values of 0.4 and 1.225 for power coefficient or the efficiency of the turbine and the air density, respectively, along with a frontal area of 10,000m². Since this is the power output for a single turbine, the first plant has been assumed to have 50 turbines, while the second has been assumed to have 100 turbines,
    \item \textbf{Solar Power Plants (2 Units):} Two solar plants generate power based on solar irradiance, contributing energy to the system. An average system efficiency of 21 percent has been assumed for both plants, followed by the first having an estimated area of 900,000m² and the second having an estimated area of 1,500,000m².
\end{itemize}

\subsubsection{Battery Storage Systems and Loads}
\begin{itemize}
    \item \textbf{Battery Units:} The system includes seven battery storage systems, each containing ten battery units. Each unit is assumed to have a capacity of 100MW.
    \item \textbf{Total Storage Capacity:}
    \begin{equation}
    \text{Total Capacity} = 7 \times 10 \times 100 \, \text{MW} = 7000 \, \text{MW}
    \end{equation}
\end{itemize}

The system is assumed to supply energy to 8 major loads, representing towns or cities with varying energy demands. 

\subsection{Region Selection and Weather Forecasting}

The regions of Rameswaram and Madurai in Tamil Nadu, known for high solar irradiation and favorable wind conditions, were chosen for testing this hybrid system. Weather forecasting is performed for the next day exclusively in order to ensure accurate results and the forecasting is updated daily with changing weather.

\section{Results}
The hybrid system was tested over one year without implementing the priority algorithm, during which significant power gaps were observed. In transitioning to a fully renewable-based grid—without constant power sources like coal or hydro plants to mitigate variability or enable priority charging—the battery storage systems struggled to meet load demands. These systems frequently reached a state of zero SoC (State of Charge), demonstrating that battery storage alone is insufficient to stabilize the hybrid renewable grid.

\begin{figure}[htbp]
\centerline{\includegraphics[scale=0.69]{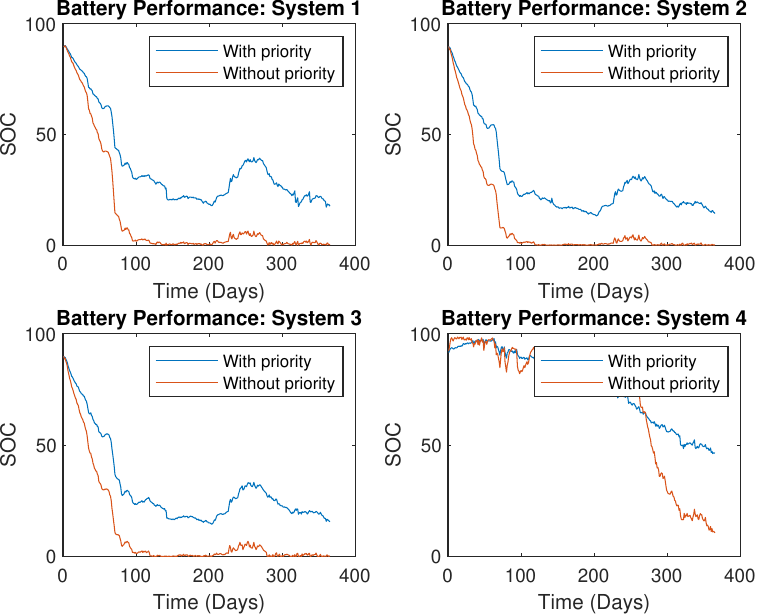}}

\end{figure}
\begin{figure}
\centerline{\includegraphics[scale=0.69]{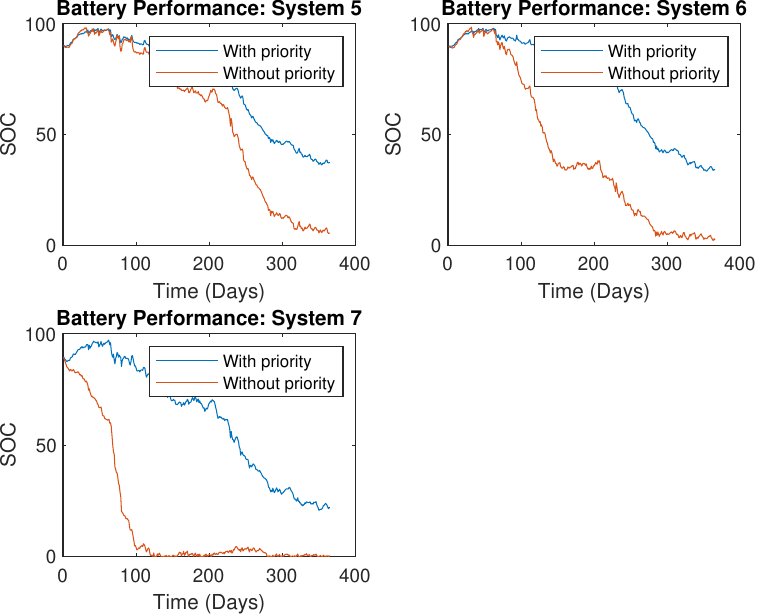}}
\caption{System stability over time with and without algorithm}
\label{fig:3}
\end{figure}
As shown in \ref{fig:3}, the fourth system slightly underperformed compared to its counterpart without priority charging. The fourth system, connected to all renewable energy sources in the experimental setup, benefits from an excess power supply compared to the other systems. In the case of equal energy distribution without prioritization, this system gains more energy and, as a result, outperforms the priority-based charging approach in this scenario.

Overall, the other six systems performed better when priority charging was enabled. None of the systems reached zero capacity throughout the year, indicating their capability to maintain grid stability under changing weather conditions.

\begin{figure}[htbp]
\centerline{\includegraphics[scale=0.65]{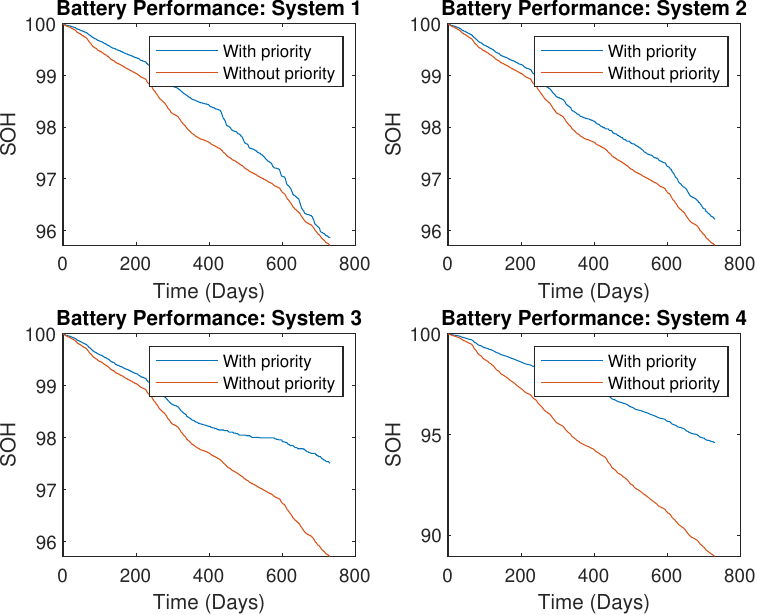}}

\centerline{\includegraphics[scale=0.65]{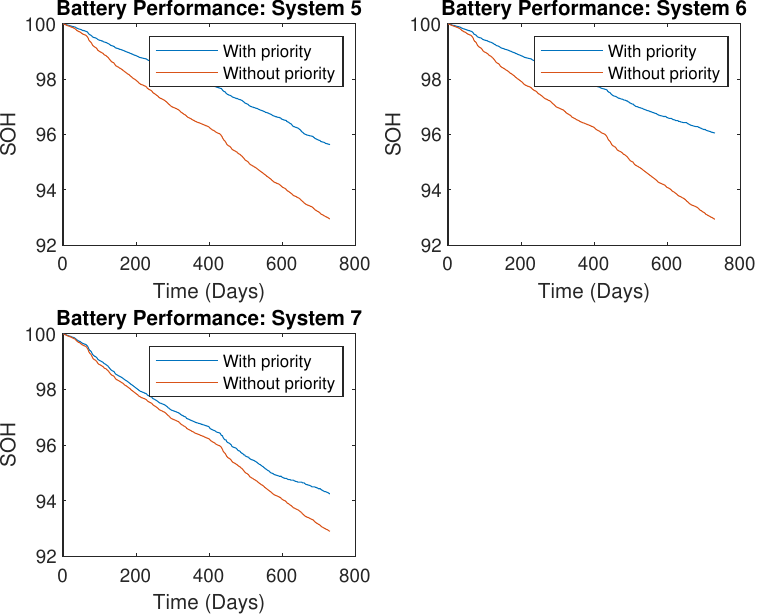}}
\caption{System degradation over time with and without the algorithm}
\label{fig:4}
\end{figure}

The health efficiency algorithm was tested for over two years using the same systems.
Battery degradation slowed across all systems after introducing the health efficiency priority algorithm. The fourth system, uniquely connected to all renewable sources, again stood out as demonstrated in \ref{fig:4}. Its counterpart without priority charging experienced the highest degradation, mainly due to overcharging with its connection to all the renewable sources and equal energy distribution. This led to the most significant observed gap in degradation between systems with and without the algorithm.

Overall, the use of the algorithm resulted in a reduction in battery degradation by 2.2\% compared to the baseline scenario. This is a significant improvement for SoH, considering eighty percent is regarded as the end of the battery's life\cite{Fan2023}. This demonstrates its effectiveness as it conserves over 1/8th of the battery's life over two years.

\FloatBarrier
\section{Conclusion}
The proposed priority charging algorithm efficiently allocates energy to the most suitable battery storage systems, addressing both short-term grid stability and long-term battery health.

Implementing a secondary health-based priority algorithm further minimizes battery degradation, extending the operational lifespan of storage systems. Simulation results show that the proposed system significantly improves energy storage efficiency, reducing the risk of battery systems reaching zero capacity and enhancing the overall resilience of the renewable energy infrastructure. 

This system offers several avenues for future improvement and expansion. One key enhancement area is the load-predicting algorithm. Instead of using SARIMA, training a custom model could give more accurate results and help plan grid management and maintenance weeks in advance.

The proposed system could be expanded to include other renewable energy sources, such as tidal energy and geothermal energy.

\section*{Acknowledgment}
We would like to thank Mars Rover Manipal, an interdisciplinary student team of MAHE, for providing the resources needed for this project.We also extend our gratitude to Dr. Ujjwal Verma for his guidance as well as Dr Soham Dutta for his support in our work.

\end{document}